\documentstyle[12pt,twoside]{article}
\begin{document}

 
\let\und=\b                     
\let\ced=\c                     
\let\du=\d                      
\let\um=\H                      
\let\sll=\l                     
\let\Sll=\L                     
\let\slo=\o                     
\let\Slo=\O                     
\let\tie=\t                     
\let\br=\u                      

 
\def\a{\alpha}
\def\b{\beta}
\def\c{\chi}
\def\d{\delta}
\def\e{\epsilon}                
\def\f{\phi}                    
\def\g{\gamma}
\def\h{\eta}
\def\i{\iota}
\def\j{\psi}
\def\k{\kappa}
\def\l{\lambda}
\def\m{\mu}
\def\n{\nu}
\def\o{\omega}
\def\p{\pi}                     
\def\q{\theta}                  
\def\r{\rho}                    
\def\s{\sigma}                  
\def\t{\tau}
\def\u{\upsilon}
\def\x{\xi}
\def\z{\zeta}
\def\D{\Delta}
\def\F{\Phi}
\def\G{\Gamma}
\def\J{\Psi}
\def\L{\Lambda}
\def\O{\Omega}
\def\P{\Pi}
\def\Q{\Theta}
\def\S{\Sigma}
\def\U{\Upsilon}
\def\X{\Xi}
 
 
\def\ca{{\cal A}}
\def\cb{{\cal B}}
\def\cc{{\cal C}}
\def\cd{{\cal D}}
\def\ce{{\cal E}}
\def\cf{{\cal F}}
\def\cg{{\cal G}}
\def\ch{{\cal H}}
\def\ci{{\cal I}}
\def\cj{{\cal J}}
\def\ck{{\cal K}}
\def\cl{{\cal L}}
\def\cm{{\cal M}}
\def\cn{{\cal N}}
\def\co{{\cal O}}
\def\cp{{\cal P}}
\def\cq{{\cal Q}}
\def\car{{\cal R}}
\def\cs{{\cal S}}
\def\ct{{\cal T}}
\def\cu{{\cal U}}
\def\cv{{\cal V}}
\def\cw{{\cal W}}
\def\cx{{\cal X}}
\def\cy{{\cal Y}}
\def\cz{{\cal Z}}

 
\def\bo{{\raise.05ex\hbox{\large$\Box$}\:}}             
\def\cbo{{\,\raise-.15ex\Sc [\,}}                       
\def\pa{\partial}                                       
\def\de{\nabla}                                         
\def\dell{\bigtriangledown}                             
\def\su{\sum}                                           
\def\pr{\prod}                                          
\def\iff{\leftrightarrow}                               
\def\conj{{\hbox{\large *}}}                            
\def\ltap{\raisebox{-.4ex}{\rlap{$\sim$}} \raisebox{.4ex}{$<$}}   
\def\gtap{\raisebox{-.4ex}{\rlap{$\sim$}} \raisebox{.4ex}{$>$}}   
\def\TH{{\raise.2ex\hbox{$\displaystyle \bigodot$}\mskip-4.7mu \llap H \;}}
\def\face{\hbox{\normalsize$\;\;\:{\raise.9ex\hbox{\oo n}\mskip-13mu \llap
        {${\buildrel{\hbox{\frtnrm ..}}\over\smile}$}}\:$}}     
\def\Face{{\raise.2ex\hbox{$\displaystyle \bigodot$}\mskip-2.2mu \llap {$\ddot
        \smile$}}}                                      
\def\dg{\sp\dagger}                                     
\def\ddg{\sp\ddagger}                                   
\def\Lhat{{\bf\rlap{\kern-.09em$\hat{\phantom L}$}L}}
\def\Lcheck{{\bf\rlap{\kern-.09em$\check{\phantom L}$}L}}
 
 
\def\sp#1{{}^{#1}}                              
\def\sb#1{{}_{#1}}                              
\def\oldsl#1{\rlap/#1}                          
\def\sl#1{\rlap{\hbox{$\mskip 1 mu /$}}#1}      
\def\Sl#1{\rlap{\hbox{$\mskip 3 mu /$}}#1}      
\def\SL#1{\rlap{\hbox{$\mskip 4.5 mu /$}}#1}    
\def\Tilde#1{\widetilde{#1}}                    
\def\Hat#1{\widehat{#1}}                        
\def\Bar#1{\overline{#1}}                       
\def\bra#1{\Big\langle #1\Big|}                 
\def\ket#1{\Big| #1\Big\rangle}                 
\def\VEV#1{\Big\langle #1\Big\rangle}           
\def\brak#1#2{\Big\langle #1\Big|#2\Big\rangle}         
\def\abs#1{\Big| #1\Big|}                       
\def\sbra#1{\left\langle #1\right|}             
\def\sket#1{\left| #1\right\rangle}             
\def\svev#1{\left\langle #1\right\rangle}       
\def\sabs#1{\left| #1\right|}                   

\def\leftrightarrowfill{$\mathsurround=0pt \mathord\leftarrow \mkern-6mu
        \cleaders\hbox{$\mkern-2mu \mathord- \mkern-2mu$}\hfill
        \mkern-6mu \mathord\rightarrow$}
\def\dvec#1{\vbox{\ialign{##\crcr
        \leftrightarrowfill\crcr\noalign{\kern-1pt\nointerlineskip}
        $\hfil\displaystyle{#1}\hfil$\crcr}}}           
\def\dt#1{{\buildrel {\hbox{\LARGE .}} \over {#1}}}     
\def\dtt#1{{\buildrel \bullet \over {#1}}}              
\def\ddt#1{{\buildrel {\hbox{\LARGE .\kern-2pt.}} \over {#1}}}
\def\der#1{{\pa \over \pa {#1}}}                
\def\fder#1{{\d \over \d {#1}}}                 
 
 
\def\frac#1#2{{\textstyle{#1\over\vphantom2\smash{\raise.20ex
        \hbox{$\scriptstyle{#2}$}}}}}                   
\def\ha{\frac12}                                        
\def\sfrac#1#2{{\vphantom1\smash{\lower.5ex\hbox{\small$#1$}}\over
        \vphantom1\smash{\raise.4ex\hbox{\small$#2$}}}} 
\def\bfrac#1#2{{\vphantom1\smash{\lower.5ex\hbox{$#1$}}\over
        \vphantom1\smash{\raise.3ex\hbox{$#2$}}}}       
\def\afrac#1#2{{\vphantom1\smash{\lower.5ex\hbox{$#1$}}\over#2}}    
\def\tder#1#2{{d #1 \over d #2 }}                 
\def\partder#1#2{{\partial #1\over\partial #2}}   
\def\brkt#1#2{{\left\langle #1 | #2 \right\rangle}} 
\def\secder#1#2#3{{\partial~2 #1\over\partial #2 \partial #3}}  
\def\on#1#2{\mathop{\null#2}\limits~{#1}}       
\def\On#1#2{{\buildrel{#1}\over{#2}}}           
\def\under#1#2{\mathop{\null#2}\limits_{#1}}    
\def\bvec#1{\on\leftarrow{#1}}                  
\def\oover#1{\on\circ{#1}}                              
 
 
\def\boxes#1{
        \newcount\num
        \num=1
        \newdimen\downsy
        \downsy=-1.64ex
        \mskip-7.8mu
        \bo
        \loop
        \ifnum\num<#1
        \llap{\raise\num\downsy\hbox{$\bo$}}
        \advance\num by1
        \repeat}
\def\boxup#1#2{\newcount\numup
        \numup=#1
        \advance\numup by-1
        \newdimen\upsy
        \upsy=.82ex
        \mskip7.8mu
        \raise\numup\upsy\hbox{$#2$}}
 
 
\newskip\humongous \humongous=0pt plus 1000pt minus 1000pt
\def\caja{\mathsurround=0pt}
\def\eqalign#1{\,\vcenter{\openup2\jot \caja
        \ialign{\strut \hfil$\displaystyle{##}$&$
        \displaystyle{{}##}$\hfil\crcr#1\crcr}}\,}
\newif\ifdtup
\def\panorama{\global\dtuptrue \openup2\jot \caja
        \everycr{\noalign{\ifdtup \global\dtupfalse
        \vskip-\lineskiplimit \vskip\normallineskiplimit
        \else \penalty\interdisplaylinepenalty \fi}}}
\def\li#1{\panorama \tabskip=\humongous                         
        \halign to\displaywidth{\hfil$\displaystyle{##}$
        \tabskip=0pt&$\displaystyle{{}##}$\hfil
        \tabskip=\humongous&\llap{$##$}\tabskip=0pt
        \crcr#1\crcr}}
\def\eqalignnotwo#1{\panorama \tabskip=\humongous
        \halign to\displaywidth{\hfil$\displaystyle{##}$
        \tabskip=0pt&$\displaystyle{{}##}$
        \tabskip=0pt&$\displaystyle{{}##}$\hfil
        \tabskip=\humongous&\llap{$##$}\tabskip=0pt
        \crcr#1\crcr}}
 
 
\def\phil{@{\extracolsep{\fill}}}
\def\unphil{@{\extracolsep{\tabcolsep}}}
 

\def\CMP{Commun. Math. Phys.}
\def\NP{Nucl. Phys. B\,}
\def\PL{Phys. Lett. B\,}
\def\PR{Phys. Rev. Lett.}
\def\PRD{Phys. Rev. D\,}
\def\CQG{Class. Quant. Grav.}
\def\IJMP{Int. J. Mod. Phys.}
\def\MPL{Mod. Phys. Lett.}

\def\ref#1{$\sp{#1]}$}
\def\Ref#1{$\sp{#1)}$}
 
 
\topmargin=.17in                        
\headheight=0in                         
\headsep=0in                    
\textheight=9in                         
\footheight=3ex                         
\footskip=4ex           
\textwidth=6in                          
\hsize=6in                              
\parindent=21pt                         
\parskip=\medskipamount                 
\lineskip=0pt                           
\abovedisplayskip=1em plus.3em minus.5em        
\belowdisplayskip=1em plus.3em minus.5em        
\abovedisplayshortskip=.5em plus.2em minus.4em  
\belowdisplayshortskip=.5em plus.2em minus.4em  
\def\baselinestretch{1.2}       
\thicklines                         
\oddsidemargin=.25in \evensidemargin=.25in      
\marginparwidth=.85in                           
 
 
\def\title#1#2#3#4{
        {\hbox to\hsize{#4 \hfill  #3}}\par
        \begin{center}\vskip.5in minus.1in {\Large\bf #1}\\[.5in minus.2in]{#2}
        \vskip1.4in minus1.2in {\bf ABSTRACT}\\[.1in]\end{center}
        \begin{quotation}\par}
\def\author#1#2{#1\\[.1in]{\it #2}\\[.1in]}

\def\AMIC{Aleksandar Mikovic\'c
\\[.1in]{\it Blackett Laboratory, Imperial College, Prince Consort Road, London
SW7 2BZ, UK}\\[.1in]}

\def\AMICIF{Aleksandar Mikovi\'c\,
\footnote{Work supported by MNTRS and Royal Society}
\\[.1in] {\it Blackett Laboratory, Imperial College, Prince Consort
Road, London SW7 2BZ, UK}\\[.1in]
and \\[.1 in]
{\it Institute of Physics, P.O. Box 57, 11001 Belgrade, Yugoslavia}
\footnote{Permanent address}\\ {\it E-mail:\, mikovic@castor.phy.bg.ac.yu}}

\def\AMSISSA{Aleksandar Mikovi\'c\,
\footnote{E-mail address: mikovic@castor.phy.bg.ac.yu}
\\[.1in] {\it SISSA-International School for Advanced Studies\\
Via Beirut 2-4, Trieste 34100, Italy}\\[.1in]
and \\[.1 in]
{\it Institute of Physics, P.O. Box 57, 11001 Belgrade, Yugoslavia}
\footnote{Permanent address}}

\def\AM{Aleksandar Mikovi\'c 
\footnote{E-mail address: mikovic@castor.phy.bg.ac.yu}
\\[.1in] {\it Institute of Physics, P.O.Box 57, Belgrade 11001, Yugoslavia}
\\[.1in]}

\def\AMsazda{Aleksandar Mikovi\'c 
\footnote{E-mail address: mikovic@castor.phy.bg.ac.yu}
and Branislav Sazdovi\'c \footnote{E-mail: sazdovic@castor.phy.bg.ac.yu}
\footnote{Work supported by MNTRS}
\\[.1in] {\it Institute of Physics, P.O.Box 57, Belgrade 11001, Yugoslavia}
\\[.1in]}

\def\AMVR{Aleksandar Mikovi\'c\,
\footnote{E-mail address: mikovic@castor.phy.bg.ac.yu}
\\[.1in] 
{\it Institute of Physics, P.O. Box 57, 11001 Belgrade, Yugoslavia}
\\[.2in]
Voja Radovanovi\'c \\[.1 in]
{\it Faculty of Physics, P.O. Box 550, 11001 Belgrade, Yugoslavia}}

\def\AMCVR{Aleksandar Mikovi\'c
\footnote{Permanent address: Institute of Physics, P.O. Box 57, 11001 
Belgrade, Yugoslavia}\footnote{E-mail: mikovic@fy.chalmers.se, 
mikovic@castor.phy.bg.ac.yu}
\\
{\it Institute of Theoretical Physics, Chalmers University of Technology,
S-412 96 Goteborg, Sweden}\\[.1in]
and
\\[.1in]
Voja Radovanovi\'c
\footnote{E-mail: rvoja@rudjer.ff.bg.ac.yu} \\
{\it Faculty of Physics, P.O. Box 550, 11001 Belgrade, Yugoslavia}}

\def\AMVVR{Aleksandar Mikovi\'c
\footnote{On leave from Institute of Physics, P.O. Box 57, 11001 
Belgrade, Yugoslavia}
\footnote{Supported by Comissi\'on Interministerial de Ciencia y Tecnologia}
\footnote{E-mail: mikovic@lie1.ific.uv.es}
\\
{\it Departamento de Fisica Te\'orica and IFIC, Centro Mixto Universidad
de Valencia-CSIC, Facultad de Fisica, Burjassot-46100, Valencia, Spain}
\\[.1in]
Voja Radovanovi\'c
\footnote{E-mail: rvoja@rudjer.ff.bg.ac.yu} \\
{\it Faculty of Physics, P.O. Box 368, 11001 Belgrade, Yugoslavia}}

\def\MBAMVVR{Maja Buri\'c
\footnote{E-mail: majab@rudjer.ff.bg.ac.yu}\\
{\it Faculty of Physics, P.O. Box 368, 11001 Belgrade, Yugoslavia}
\\[.1in]
Aleksandar Mikovi\'c
\footnote{On leave from Institute of Physics, P.O. Box 57, 11001 
Belgrade, Yugoslavia}
\footnote{Supported by Comissi\'on Interministerial de Ciencia y Tecnologia}
\footnote{E-mail: mikovic@lie1.ific.uv.es}
\\
{\it Departamento de Fisica Te\'orica and IFIC, Centro Mixto Universidad
de Valencia-CSIC, Facultad de Fisica, Burjassot-46100, Valencia, Spain}
\\[.1in]
Voja Radovanovi\'c
\footnote{E-mail: rvoja@rudjer.ff.bg.ac.yu} \\
{\it Faculty of Physics, P.O. Box 368, 11001 Belgrade, Yugoslavia}}

\def\AMV{Aleksandar Mikovi\'c
\footnote{On leave from Institute of Physics, P.O. Box 57, 11001 
Belgrade, Yugoslavia}
\footnote{Supported by Comissi\'on Interministerial de Ciencia y Tecnologia}
\footnote{E-mail: mikovic@lie1.ific.uv.es}
\\
{\it Departamento de Fisica Te\'orica and IFIC, Centro Mixto Universidad
de Valencia-CSIC, Facultad de Fisica, Burjassot-46100, Valencia, Spain}}

\def\endtitle{\par\end{quotation}\vskip3.5in minus2.3in\newpage}
 
 
\def\endabstract{\par\end{quotation}
        \renewcommand{\baselinestretch}{1.2}\small\normalsize}
 
 
\def\xpar{\par}                                         

\def\letterhead{
        \centerline{\large\sf INSTITUTE OF PHYSICS}
        \centerline{\sf P.O.Box 57, 11001 Belgrade, Yugoslavia}
        \rightline{\scriptsize\sf Dr Aleksandar Mikovi\'c}
        \vskip-.07in
        \rightline{\scriptsize\sf Tel: 11 615 575}
        \vskip-.07in
        \rightline{\scriptsize\sf E-mail: MIKOVIC@CASTOR.PHY.BG.AC.YU}}

\def\sig#1{{\leftskip=3.75in\parindent=0in\goodbreak\bigskip{Sincerely yours,}
\nobreak\vskip .7in{#1}\par}}

\def\ssig#1{{\leftskip=3.75in\parindent=0in\goodbreak\bigskip{}
\nobreak\vskip .7in{#1}\par}}

 
\def\ree#1#2#3{
        \hfuzz=35pt\hsize=5.5in\textwidth=5.5in
        \begin{document}
        \ttraggedright
        \par
        \noindent Referee report on Manuscript \##1\\
        Title: #2\\
        Authors: #3}
 
 
\def\start#1{\pagestyle{myheadings}\begin{document}\thispagestyle{myheadings}
        \setcounter{page}{#1}}
 
 
\catcode`@=11
 
\def\ps@myheadings{\def\@oddhead{\hbox{}\footnotesize\bf\rightmark \hfil
        \thepage}\def\@oddfoot{}\def\@evenhead{\footnotesize\bf
        \thepage\hfil\leftmark\hbox{}}\def\@evenfoot{}
        \def\sectionmark##1{}\def\subsectionmark##1{}
        \topmargin=-.35in\headheight=.17in\headsep=.35in}
\def\ps@acidheadings{\def\@oddhead{\hbox{}\rightmark\hbox{}}
        \def\@oddfoot{\rm\hfil\thepage\hfil}
        \def\@evenhead{\hbox{}\leftmark\hbox{}}\let\@evenfoot\@oddfoot
        \def\sectionmark##1{}\def\subsectionmark##1{}
        \topmargin=-.35in\headheight=.17in\headsep=.35in}
 
\catcode`@=12
 
\def\sect#1{\bigskip\medskip\goodbreak\noindent{\large\bf{#1}}\par\nobreak
        \medskip\markright{#1}}
\def\chsc#1#2{\phantom m\vskip.5in\noindent{\LARGE\bf{#1}}\par\vskip.75in
        \noindent{\large\bf{#2}}\par\medskip\markboth{#1}{#2}}
\def\Chsc#1#2#3#4{\phantom m\vskip.5in\noindent\halign{\LARGE\bf##&
        \LARGE\bf##\hfil\cr{#1}&{#2}\cr\noalign{\vskip8pt}&{#3}\cr}\par\vskip
        .75in\noindent{\large\bf{#4}}\par\medskip\markboth{{#1}{#2}{#3}}{#4}}
\def\chap#1{\phantom m\vskip.5in\noindent{\LARGE\bf{#1}}\par\vskip.75in
        \markboth{#1}{#1}}
\def\refs{\bigskip\medskip\goodbreak\noindent{\large\bf{REFERENCES}}\par
        \nobreak\bigskip\markboth{REFERENCES}{REFERENCES}
        \frenchspacing \parskip=0pt \renewcommand{\baselinestretch}{1}\small}
\def\unrefs{\normalsize \nonfrenchspacing \parskip=medskipamount}
\def\Item{\par\hang\textindent}
\def\Itemitem{\par\indent \hangindent2\parindent \textindent}
\def\makelabel#1{\hfil #1}
\def\topic{\par\noindent \hangafter1 \hangindent20pt}
\def\Topic{\par\noindent \hangafter1 \hangindent60pt}

\def\Sch{Schwarzschild }
\def\MBVRAMV{Maja Buri\'c
\footnote{E-mail: majab@rudjer.ff.bg.ac.yu} and
 Voja Radovanovi\'c\footnote{E-mail: rvoja@rudjer.ff.bg.ac.yu}\\
{\it Faculty of Physics, P.O. Box 368, 11001 Belgrade, Yugoslavia}
\\[.1in]
Aleksandar Mikovi\'c\footnote{E-mail: mikovic@lie.ific.uv.es}\\
{\it Departamento de Fisica Te\'orica and IFIC, Centro Mixto Universidad
de Valencia-CSIC, Facultad de Fisica, Burjassot-46100, Valencia, Spain
and Institute of Physics, P.O. Box 57, 11001 Belgrade, Yugoslavia}}

\title{One-loop corrections for Schwarzschild black hole via 2D dilaton
gravity}
{\MBVRAMV}{}{September 1998}

\noindent We study 
quantum corrections for the Schwarzshild black hole by considering it
as a vacuum solution of a 2D dilaton gravity theory obtained by 
spherical reduction of 4D gravity coupled with matter. 
We find perturbatively the vacuum solution for the standard one-loop effective
action
in the case of null-dust matter and in the case of minimally 
coupled scalar field. The corresponding state is in both cases 2D 
Hartle-Hawking vacuum, and we evaluate the corresponding
quantum corrections for the thermodynamic parameters of the
black hole. We also find that the standard effective action does not 
allow boundary conditions corresponding to a 4D Hartle-Hawking vacuum state. 

\endtitle

\sect{1. Introduction}

One of the most interesting problems in quantum gravity is the
Hawking radiation of black holes \cite{h}. As we do 
not yet have a complete quantum theory of gravity, the full description of
this phenomenon is still missing. Note that the string theory has 
given a microscopic explanation  of this process \cite{cm}. However, a complete
formalism for calculating the large-radius backreaction effects does
not exist. These effects are described by the effective action,
 and in the absence of a 
complete formalism for calculating the effective action, one has to 
resort to various approximations.

One way is to quantize the matter fields in
the fixed black hole background. It is then  possible to calculate the
quantum corrections to the classical metrics, the
spectrum of the radiation, temperature, etc. The back-reaction of the
radiation to the metric is calculated by defining the
appropriate expected value of the energy-momentum tensor of the matter
field \cite{bd,y,ahwy} and solving the corresponding "one-loop" equation
$$ R_{\m\n}-g_{\m\n}R/2=\svev{T_{\m\n}}\quad.\eqno(1.1)$$ 
This approach brought a fairly clear qualitative
picture of the process. A better approach is to 
integrate the gravitational and the matter field  in the functional integral 
and obtain the one-loop
effective action, which would allow a  background independent approach. 
This, unfortunately, cannot be done in four spacetime
dimensions (4D) because of the nonrenormalizability of gravity. However, in 
two spacetime dimensions (2D), gravity is renormalisable, and this procedure 
can be done.  Therefore if one considers the spherically symmetric general
relativity with matter as a 2D field theory, then it is possible to calculate 
the corresponding one-loop
effective action. It is plausible to assume that such an action would be a good
approximation for a full theory for large radius. 
Recently several papers have appeared on how
to calculate this effective action \cite{bh,no,klv,mr4}. All these papers 
gave similar results, modulo ambiguities in coefficients of
certain counterterms.

Given this action, one can now start to investigate the one-loop backreaction
effects. The simplest thing is to investigate the static vacuum solutions.
This approach has been already started in \cite{fis}
where the quantum corrections to the Reissner-Nordstrom black
hole and the corresponding thermodynamic properties were
calculated. In that paper the authors have used only the
Polyakov-Liouville term in the effective action, while it is known
that the large radius null-dust action also contains additional
local terms \cite{mr}. It is very well known fact \cite{rst,bpp} that the 
local terms
in the effective action can influence  the form of the solution. This
is one reason why we consider the null-dust model. Another reason is that
the null dust model can serve as a preparatory study for the more complicated,
and more realistic model which is spherically symmetric reduction of
general relativity with minimally coupled scalar field (SSG). 
The one-loop effective
action for SSG model is qualitatively different from the null-dust action.
There is a new nonlocal term due to the
coupling between the dilaton and the matter field. Its influence on
the form of the one-loop solution could be important, so that we  
calculate the correction
to the Schwarzschild black-hole solution and the corresponding
corrections to the thermodynamic parameters.
As we mentioned, there exists some ambiguity in the literature
about the $R\F$ coefficient in the effective action for SSG, and
therefore we investigate how its value 
affects the physical parameters of the solution.

The third motivation is to compare the properties of this solution to those 
obtained in 4D via (1.1) \cite{ahwy}, in order to see how good is the 
2D effective action approach.

The plan of the paper is the following: in Sect. 2 we briefly review
the spherically symmetric one-loop effective action and transform it to a
local form by using two auxiliary fields which mimic trace anomaly in
 our case.    
As a warm-up exercise we consider first a simpler model of null-dust 
matter in Sect. 3: its equations of
motion, perturbative solutions, corrections to the radius, temperature
and entropy of the black hole. This section is very similar to the
\cite{fis}, but we consider only uncharged black hole. In Sect. 4 
the same is done for the
spherically symmetric model, and we present the results for an arbitrary $R\F$
coefficient. In section 5 we present our conclusions. Appendix A contains
all relevant formulae given for the action with arbitrary
$R\Phi$ coefficient. The alternative calculation of
entropy  using  the conical singularity method is given in Appendix B.

\sect{2. One-loop effective action}

Spherically symmetric reduction of
the Einstein-Hilbert action in 4D gives the following 2D dilaton
gravity action
$$ \G_0 = {1\over 4G}  \int d^2 x \sqrt{-g} e^{-2\F}\left( R
 + 2 ( \nabla \F)^2 +2e^{2\F} \right) \quad ,
\eqno(2.1)$$
where $G,\F, g_{\m\n}$ are the Newton constant, dilaton and two
dimensional metric, respectively. The 4D line element is given by
$$ {ds_{(4)}}\sp 2 = g_{\m\n} dx\sp\m dx\sp\n + e\sp{-2\F}d\O\sp 2 
\eqno(2.2)$$
so that $r =r_0 e\sp{-\F}$ can be identified as the spatial radius in
appropriate gauge ($r_0$ is an arbitrary length constant, which is needed
for dimensional reasons).

If one couples minimally $N$ scalar 2D fields $f_i$ to this action, one gets 
the
null-dust model, with the action
 $$ \G_0 =  {1\over 4G}\int d^2 x \sqrt{- g}\left[ e^{-2\F}\left(  R
 + 2 (  \nabla \F)^2 + 2e^{2\F} \right) - 2G\su_{i=1}^N
(  \nabla f_i )^2 \right]\quad ,\eqno(2.3)$$
where the number of scalar fields $N$ is introduced 
in order to obtain the
semiclassical approximation from the large-$N$ limit.
If  one adds $N$ scalar fields which are minimally
coupled to gravity in 4D and afterwards performs the spherically symmetric
reduction the action  becomes
 $$ \G_0 =  {1\over 4G}\int d^2 x \sqrt{- g}\left[ e^{-2\F}\left(  R
 + 2 (  \nabla \F)^2 + 2e^{2\F} \right) - 2G\su_{i=1}^N
e^{-2\F}(  \nabla f_i )^2 \right]\, .\eqno(2.4)$$
The null-dust model differs from SSG by the absence of  coupling
between scalar field and  dilaton. 
It can be thought of as a large radius 
approximation
to SSG, since the rescaling $f_i \to f_i /r$ in (2.4) will give (2.3) plus
terms of order $1/r$. 

The one-loop effective action for the model (2.3) has been found in
\cite{mr}. The large-N, large-$r$ one-loop part of the action is given by
$$\li{ \G =& {1\over 4G}\int d^2 x \sqrt{- g}\left[ e^{-2\F}\left(  R
 + 2 (  \nabla \F)^2 + 2e^{2\F} \right) - {1\over 2}\su_{i}
(  \nabla f_i )^2 \right]\cr
& - 
{N\over 96\p } \int d^2x \sqrt {-g} \left[
R{1\over \Box } R- 
(\nabla \F )^2+ 2 R\F \right]
\ .&(2.5)\cr}$$
Note that this action differs from the one used in \cite{fis}
by the presence of two local terms, $(\nabla \F )^2$ and $ 2 R\F$. The
action (2.5) is very similar to the BPP model \cite{bpp}. The action (2.5) can
be rewritten as
$$\li {\G = &{1\over 4G} {\Big(} \int d^2x\sqrt{- g}[ r^2R+2(\nabla r)^2
+2-2G\sum _i(\nabla f_i )^2]-\cr
 & -\kappa\int
d^2x\sqrt{- g}[ R {1\over\Box}R-2R\log r-{(\nabla r)^2\over r^2}]
{\Big)}\ , &(2.6)\cr}$$
where $r= r_0 e^{-\F}$ and  $\k ={NG\over 24\p }.$
Since we are interested in the  vacuum
solutions, we will take $f_i =0$.
The action (2.6) can be transformed into a local form by using 
an auxiliary scalar field 
$\psi$:
$$\li{\G ={1\over 4G}{\Big(} &\int d^2x\sqrt {-g}[ r^2R+2(\nabla r)^2+2]
\cr
 &-\kappa\int
d^2x\sqrt{- g}[2R\psi+(\nabla\psi )^2-2R\log r-{(\nabla r)^2\over
r^2}]{\Big)}\ ,
&(2.7)\cr}$$
 where $\psi$ satisfies
$$\Box\psi=R\ .$$

In the SSG case
the quantization of the matter fields gives the following one-loop
correction to the effective action \cite{bh,no,klv,mr4},
$$\G_1= -{N\over 96\p }\int d\sp 2 x \sqrt{-g} \left( R{1\over \Box }R
-12 R{1\over \Box}(\nabla \F )^2  + c R\F
 \right) \ ,\eqno(2.8)$$
where $c$ is a non-zero constant, whose numerical value depends on the
quantization procedure used. The dimensional regularization method gives
$c=12$ \cite{no,klv}. Dimensional regularization with a rescaled metric 
also gives $c=12$, but after returning to the original variables  
$c$ changes to $c=14$ \cite{mr4}. The 
$\zeta$-function regularization gives $c= -4$ \cite{bh}.
It will turn out that this ambiguity does not change any of our calculations
essentially, so we will
proceed with the value $c=12$ and summarize the results for the
case of an arbitrary coefficient in the Appendix A.

Note that one can also calculate the one-loop contributions due to all fields
\cite{mr4}. This calculation simplifies when a rescaled metric
$\tilde g_{\mu\nu}=e^\F g_{\mu\nu}$ is quantized,
so that in the large-$N$ and  large-$r$ limit one obtains \cite{mr4}  
$$\tilde\G_1= -{N\over 96\p }\int d\sp 2 x \sqrt{-\tilde g} \left( \tilde R
{1\over \Box }\tilde R
-12 \tilde R{1\over \Box}(\tilde \nabla \F )^2  + 12\tilde R\F
+ 12 \su_i \tilde R{1\over \Box }(\tilde \nabla f_i)^2 \right) \, .\eqno(2.9)$$
When compared to (2.8), the last term in (2.9) is new, 
and it comes from the finite parts of the graviton-dilaton loops.
Since we are interested in vacuum solutions, for which $f_i =0$,
this difference will not be essential. However, when transforming back
to the original metric, the $f_i=0$ limit of (2.9)
will produce an additional  term in (2.8), proportional to 
$\int\sqrt{-g}(\nabla\F)\sp 2$, which is
of the relevant order in $1/r$. Since it is not clear whether the quantization
of the original metric will produce this term, in this paper we will consider
only the quantum corrections given by (2.8), although one should keep in mind
that additional counterterms are possible due to different quantization 
procedures.

The analysis of the action given by the correction (2.8) will simplify if it
is written in the local
form. Note that in this case we are dealing with two
nonlocal terms, $ R{1\over \Box }R$ and $ R{1\over \Box }(\nabla \F)^2$.
This implies that we have to introduce two auxiliary fields, $\psi$
and $\chi$. After a tedious calculation we get the following local form of the
 correction 
$$\G_1=-{N\over 96\p}\int d^2x\sqrt {-g} \left[2R(\psi -6\chi )+(\nabla
\psi)^2-12(\nabla \psi )(\nabla \chi)-12\psi (\nabla \F )^2+12R\F \right].
\eqno(2.10)$$
The additional fields then satisfy the equations of motion
$$\Box\psi =   R\ , \eqno(2.11)$$
and
$$\Box\chi =(\nabla \F)^2\ . \eqno(2.12)$$

Note that it is now easy to obtain the expression for trace anomaly for the SSG
case from (2.10) and
(2.11-12). It is given by
$$T={1\over 24\p}(R-6(\nabla \F)^2+6\Box \F). \eqno(2.13)$$
In the  case of the null-dust model $T$ is given by
$$T={1\over 24\p}R , \eqno(2.14)$$
if we take the Polyakov-Liouville term only, or
$$T={1\over 24\p}(R+\Box \F) \eqno(2.15)$$
in the case (2.7). 

\sect{3. Spherically symmetric null-dust model}

We will discuss first the simpler model of null-dust matter in order to
prepare for the more complicated case of spherically symmetric scalar field. 
We will use the action in the form
$$\li{\G ={1\over 4G}{\Big(} &\int d^2x\sqrt{- g}[ r^2R+2(\nabla r)^2+2]
\cr
 &-\kappa\int
d^2x\sqrt{- g}[2R\psi+(\nabla\psi )^2-2R\log r-{(\nabla r)^2\over
r^2}]{\Big)}.
&(3.1)\cr}$$
The variation of this action with respect
to the fields $\psi$, $r$ and $g^{\m\n}$ gives the equations of motion:
$$\Box\psi =R\ \ ,\eqno(3.2)$$
$$rR-2\nabla ^2 r=-\kappa ({R\over r}+{(\nabla r)^2\over r^3}-{\nabla ^2r\over
r^2})\ \ ,\eqno(3.3)$$ and
$$\li {-2r\nabla _{\m}\nabla_{\n}r &+g_{\m\n}(\nabla ^2 r^2-(\nabla r)^2-1)=
-\k T_{\mu\nu}=\cr
&=\k(-2\nabla _{\m}\nabla _{\n}\psi +\nabla _{\m}\psi\nabla _{\n}\psi
+g_{\m\n}(2R-{1\over 2}(\nabla \psi )^2)+ \cr
&+2{\nabla
_{\m}\nabla_{\n}r\over r}-3{\nabla _{\m}r\nabla _{\n}r\over r^2}+g_{\m\n}(-2{
\nabla ^2 r\over r}+{5\over 2} {(\nabla r)^2\over r^2}))\ . &(3.4)\cr}$$

From (3.3-4) we can obtain the expression for the scalar curvature:
$$R={2-2(\nabla r)^2+{\k \over r^2}\over r^2}\quad.\eqno(3.5)$$

This expression shows that the singularity of the curvature
is at $r=0$, as in the classical case. The solution of the equation
(3.3) for the field $r$ is $r=x^1$, so the field $r$ really
 has the meaning of the radial
coordinate, in accordance with (2.2). In \cite{fis}, the quantum
correction shifts the space-time
singularity to the value $r_{cr}^2=2\k $; in our case it stays at $r=0$.
This is a consequence of
the fact that the quantum correction in \cite{fis} is given by the
Polyakov-Liouville term only, while in our case there are  two additional 
local terms.  

The classical solution (corresponding to  $\k =0$) of the equations
of motion (3.3-4) is 
$$r=x^1, \ \ ds^2=-f_0(r) dt^2+{1\over f_0(r)}dr^2\quad, \eqno(3.6) $$ 
 where $t=x^0$, $f_0=1-{2GM\over r}=1-{a\over r}.$
  We are interested in the quantum correction
of this solution in the semiclassical approximation.
This means that we are searching for the perturbative solution
of the equations (3.2-4) by taking $\kappa$  as a small parameter. 
By introducing the static ansatz as in \cite{fis}:
$$ds^2=-f(r)e^{2\f (r)} dt^2+{1\over f(r)}dr^2\ 
,\eqno(3.7) $$
with
$f(r)=1-{a\over r}+\k {m(r)\over r},$ 
 we get, in the first order in $\k$:
 $$\psi ^{\prime}=-{f_0^{\prime}\over f_0}+{C\over f_0}\ .
\eqno(3.8)$$
$$\f^{\prime}={\k \over 2rf_0}\left(2f_0^{\prime\prime}-{f^{\prime 2}\over
f_0}+{C^2\over
f_0}-{3f_0\over r^2}\right)\ , \eqno(3.9)$$
$$m^{\prime}= \left(-2f_0^{\prime\prime}+{1\over 2f_0}(f_0^{\prime
2}-C^2)-{f_0^\prime\over r}+{5\over 2}{f_0\over
r^2}\right)\ . \eqno(3.10)$$
The integration constant $C$ can be determined from the condition that
the behaviour of $T_{00}$ at infinity is thermal. This boundary condition
follows from the fact that the static solution we are constructing describes
a black hole in thermal equilibrium with the Hawking radiation (the black
hole emits as much energy as it absorbs, so that its mass does not change). 
Since the matter in this model behaves as  a free 2D scalar field,
we take that $T_{00}$ at infinity (in the zero-th order in $\k$)
has the 
temperature dependence of a 1D free bose-gas $(\pi /6) T_H^2$, where 
$T_H$ is the classical Hawking temperature $4\p T_H={1\over a}$.  
This is consistent with the fact that for spherical null-dust
$\svev{T_{uu}} = \svev{ T_{vv}} = 1/48\p (4M)^2$ in the 
Hartle-Hawking vacuum, where $u$ and $v$ are
the asymptotically flat Schwarzschild coordinates \cite{cmn}.
This gives $C^2={1\over a^2}$.
By integrating (3.9-10) we obtain 
$$m(r)=-{1\over 2a}\log ({r\over l})-{r\over 2a^2} -{2\over r}\ ,
\eqno(3.11)$$
$$\f (r)= F(r)-F(L)\quad,$$
$$F(r)=\k \left({1\over 2a^2}\log ({r\over l})-{1\over ar}\right)\ ,
\eqno(3.12)$$
where $L$ and $l$ are the integration constants. 
We have assumed that our system is in a 1D spatial box of size $L (a\ll L)$,
and that  $e^{2\f}=1$ as $r\to L$. 

The position of the horizon $r_h$ can be
found perturbatively from the condition $f(r_h)=0$:
$$r_h=a-\k m(a)=a+\k \left({1\over 2a}\log ({a\over l})+{5\over
2a}\right)\quad. \eqno(3.13)$$
The Hawking temperature  also changes due to the quantum corrections. In
the case of the metric of the form (3.7), it is given by (see Appendix
B)
$$4\p T_H=e^\f f^{\prime }|_{r=r_h}\ ,\eqno(3.14)$$
which, after insertion  of $\f$ and $m$, gives
$$4\p T_H={1\over a}\left(1-\k ({5\over 2a^2}+F(L))\right)\ .\eqno(3.15)$$
The second term in (3.15) is the quantum correction of the temperature.

In order to calculate the entropy of the quantum corrected black hole
solution, we will use  Wald
technique \cite{w}. In refs \cite{m,jkm,iw} it was shown
that for the lagrangian of the form $L=L(f_{m}, \nabla f_{m},
g_{\m\n} , R_{\m\n\r\s})$ ($f_{m}$ are the matter fields), the
entropy is given by
$$S=-{2\p}\e _{\a\b}\e_{\c\d}{\pa L\over \pa R_{\a\b\c\d}}\quad,\eqno(3.16)$$
evaluated at the horizon. Hence the evaluation of the entropy via Wald's method
does not require the knowledge of the boundary terms (which might occur
in the action), which are necessary
if one evaluates the entropy from the Euclidean action \cite{fis}.
For the lagrangian given by 
(3.1), we obtain
$$S={\p a^2\over G}+{\k \p \over G}\left(3+\log {a\over l}\right)\quad .
\eqno(3.17)$$
The first term in (3.17) is the
Bekenstein-Hawking entropy, while the second term is the quantum
correction. The same result is obtained when we define properly the boundary
terms for the action (3.1), and  use the conical singularity method
and this calculation is presented in Appendix B.

Using equations (3.15) and (3.17) and the first law of thermodynamics 
$TdS=dE$
(where the local temperature is given by $T={T_H\over \sqrt{-g_{00}(L)}}$), for
the energy of the system we get the following expression:
$$E=M+{\k\over 4G^2M}\left({7\over 4}+{1\over 2}\log{L\over l}\right) \, .
\eqno(3.18)$$

The results which we have obtained are 
qualitatively similar to those of
\cite{fis}. The difference in the numerical coefficients is due to the
extra terms in the one-loop effective action. 
We also have a logarithmic correction to
the entropy (term proportional to $\log M$). The main
difference is that in our model the singularity of the curvature
stays at the origin $r=0$.  

\sect{4. Spherically symmetric scalar field model}

We now examine the more realistic, and more complicated case of
spherically symmetric scalar field.
By adding the actions (2.4) and (2.8) and by setting $f_i =0$ we get 
the one-loop effective action
$$\li{\G ={1\over 4G}{\Big(}& \int d^2x\sqrt{- g}[ r^2R+2(\nabla r)^2+2]-
\cr
& -\kappa\int
d^2x\sqrt{- g}[ R{1\over\Box}R-12R{1\over \Box}
{(\nabla r)^2\over r^2}-12R\log r] {\Big)}\ , &(4.1)\cr} $$
which is, in the local form,
$$\li{\G ={1\over 4G}{\Big(} \int d^2x\sqrt{- g}&[ 
r^2R+2(\nabla r)^2+2]-\cr
 -\kappa\int d^2x \sqrt{- g} &[ 2R(\psi -6\chi )+(\nabla \psi )^2-\cr
 &-12 (\nabla \psi
)(\nabla \chi) -12{\psi (\nabla r)^2\over r^2}-12R\log r]{\Big)} \ .
&(4.2)\cr } $$

Varying the action (4.2) with respect to $r$, $g_{\m \n}$, $\chi$ and $\psi$
we obtain the following equations of motion
$$ -rR+2\nabla ^2r =6\k \left(2{\psi (\nabla r)^2\over r^3}-2 {\psi \nabla
^2r \over
r^2}-2{(\nabla \psi)(\nabla r)\over r^2} +{R\over r}\right)\ ,
\eqno(4.3)$$
$$\li{-2r&\nabla _{\m}\nabla_{\n}r+g_{\m\n}(\nabla ^2 r^2-(\nabla r)^2-1)=\cr
=\k&\left(g_{\m\n}(2\Box\psi-12\Box\chi -12{\nabla ^2 r \over
r}+(12+6\psi){(\nabla r)^2\over r^2}-{1\over 2}(\nabla \psi)^2+6(\nabla
\psi)(\nabla \chi))+\right.\cr
&+\nabla _{\m}
\psi \nabla _{\n}\psi-12\nabla _{\m}
\psi \nabla _{\n}\chi  -2\nabla _{\m} \nabla _{\n}\psi +\cr
&\left.+12\nabla _{\m}
 \nabla _{\n}\chi +12{\nabla _{\m}\nabla _{\n}r\over r}-12{\nabla _{\m}
r \nabla _{\n}r\over r^2}(\psi +1)\right)\quad, &(4.4)\cr}$$
$$\Box\psi =R\ , \eqno(4.5)$$

$$\Box\chi ={(\nabla r)^2\over r^2}\ . \eqno(4.6)$$
 
 The vacuum solution
of the classical ($\k =0$) equations is the Schwarzschild black hole (3.6).
We will, again, solve (4.3-6) perturbatively in $\k$,
starting with the static ansatz (3.7).
Integration of the  equations for $\psi$ and $\chi$ in the zero'th order 
in $\k$ gives
$$\psi = Cr +Ca\log{ r-a\over l}- \log {r-a \over r}\quad ,\eqno (4.7)$$
$$\chi ^\prime ={2Dr^2-2r+a\over 2r(r-a)}\quad ,\eqno(4.8)$$
where $C,D$ and $l$ are the integration constants. 

The 00 and 11 components of the equation (4.4), to the first order in $\k$, are
$$\li{-rf^{\prime}-f+1&=\k \tau ^0_0=\cr
&=\k {\Big(} f^{\prime}\psi ^{\prime}-2R
-6f^{\prime}\chi^{\prime}+{1\over 2}f\psi^{\prime 2}-6f\psi
^{\prime}\chi^{\prime} -6{f\psi \over r^2}+6{f^{\prime}\over r}{\Big)} 
\ , &(4.9)\cr}$$
$$\li{&rf^{\prime}+2\f^{\prime}fr+f-1=\k \tau ^1_1=
\k {\Big(} -f^{\prime}\psi ^{\prime}
+2R+\cr
& +6f^{\prime}\chi^{\prime}+{1\over 2}f\psi^{\prime 2}-6f\psi
^{\prime}\chi^{\prime} -6(\psi +2){f \over r^2}-6{f^{\prime}\over
r}-2f\psi ^{\prime\prime}+12f\chi ^{\prime\prime}{\Big)}
\ . &(4.10)\cr} $$
From (4.9-10) we easily obtain 
$$m^{\prime}=\tau _0^0,\ \ \f ^{\prime}={\k \over 2rf}(\tau ^1_1 +
\tau _0^0)\quad. \eqno(4.11)$$
Using the expressions for $\psi $ and $\chi ^{\prime}$ (4.7-8), for
$m$ and $\f$ we obtain the solutions
$$\li{m(r)=&{11a\over 4r^2}+{-5+6aC\over 2r}-{1\over
2}(C^2-12CD)r -\cr
&-{5\over 2a}\log {r-a\over r}+{6\over r}\log  {r-a\over r}-{3a\over r^2}\log {r-a\over r}-\cr
&-{a(C^2-12CD)\over 2}\log {r-a\over l}
-{6aC\over r}\log {r-a\over l}+{3a^2C\over r^2}\log {r-a\over l}
\ , &(4.12)\cr}
$$
and
$$\f  (r)=\k (F(r)-F(L))\quad,\eqno(4.13)$$
 where $F(r)$ is given by
$$\li{F(r)=&{3\over 4r^2}+{2+3aC\over
ar}+{1-6aC-a^2(C^2-12CD)\over 2a(r-a)}+ \cr
&+{5\over 2a^2}\log {r-a\over r}-{3\over r^2}\log {r-a\over r} +\cr
&+{(C^2-12CD)\over 2}\log {r-a\over l}+{3
aC\over r^2}\log {r-a\over l}
\quad. &(4.14)\cr}
$$
The constant $L$ is chosen as a regularization parameter for large radius $r$.
As the integration constant in $m(r)$ can be absorbed in the definition of the
mass of the black hole \cite{y,ahwy}, in (4.12) we have fixed it by using the
previously introduced length scale $l$ only.

In order to fix the values of the constants $C$ and $D$, let us analyze
the properties of the solution (4.12-14). Let us first calculate the
determinant of the metric tensor, $g=e^{2\f}$. In the first order in
$\k$ we get
$$g=1+2\f =1+2\k (F(r)-F(L))\ ,$$
which, in the vicinity of the point $r=a$ reduces to
$$g={\rm const} +\k {1-6aC-a^2(C^2-12CD)\over a(r-a)}
-\k {1-6aC-a^2(C^2-12CD)\over a^2}\log{(r-a)}\ .$$
Obviously $g$ is divergent at $r=a$ unless 
 $$1-6aC-a^2(C^2-12CD)=0\quad.\eqno(4.15)$$

The position of the horizon is determined from the condition 
$f(r_h)=0$. If we assume that the correction is perturbative,
$r_h=a+\k r_1$, we get $$r_1=-m(a+\k r_1)\quad, \eqno(4.16)$$ 
which gives
$$\li{r_1 =&-{1+12aC\over 4a}+{1\over
2}a(C^2-12CD) +\cr &+{1\over 2a}\log a
-{1-6aC-a^2(C^2-12CD)\over 2a}\log (\k r_1)+\cr
&-3C\log l-{a(C^2-12CD)\over 2}\log l
\ . &(4.17)\cr}
$$  
Note that the term proportional to $\log (\k r_1)$,
 which appears in this formula, is
divergent in the limit $\k\to 0$, so one should take that
 the corresponding coefficient vanishes:
$1-6aC-a^2(C^2-12CD)=0$, in order to stay in the region of the
perturbative calculation.

The third argument also forces us to fix $C$ and $D$ in accordance with
(4.15). Namely, if we calculate the 
Hawking temperature for the one-loop
corrected geometry using  (3.14),
 we get
$$4\p T_H={1\over r_h}(1+\k m^\prime (r_h) )e^{\f (r_h)}\ .\eqno(4.18)$$
Using the expressions for $m$, $\f$ and $r_h$, 
$$\li{4\p T_H=&{1\over a}\left( 1+{1-6aC-a^2(C^2-12CD)\over ar_1}
\right.+\cr
&+\k\left( -{1\over 2a^2}+{3C\over a}-(C^2-12CD)+{C^2-12CD\over 2}\log l
-F(L)\right) 
 .&(4.19)\cr}$$
 Here, also,
there is a nonperturbative term proportional to the
factor $1-6aC-a^2(C^2-12CD)$. Therefore, (4.15) fixes one condition for
$C$ and $D$.

If we calculate the scalar curvature $R$ in the first order in $\k$ we obtain
 $$\li{R&={2a\over r^3}+{\kappa\over 2ar^5}\Big( 
 a[3a(12Cr-25)-36a^2C\log l+\cr
&+2r(5+(C^2-12CD)r^2)]+\cr
&+2(18a^2-5r^2)\log r+\cr
&+2[18a^2(aC-1)+(5+(C^2-12CD)a^2)r^2]\log(r-a)\Big) .&(4.20)\cr}$$ 
 From the last expression we see that the curvature singularities are 
  $r=0$ and $r=a$. In order to remove the second singularity and keep
the calculation perturbative, we get the second relation between $C$
and $D$:
$$ 18aC-13+(C^2-12CD)a^2=0\ .\eqno(4.21)$$
The conditions of applicability of perturbative
calculation (4.15) and (4.21) give  
$$C={1\over a},\ \ D={1\over 2a}\eqno(4.22)$$

We will now analyse the behaviour of the stress-energy tensor
for the SSG case. Using (4.12-14), we get 
$$\li{T_{00}&={1\over96\p r^4}{\Big(} -5a^2+
4ar-12a^2Cr+18aCr^2+(C^2-12CD)r^4+\cr
&+12(a-r)^2(\log {r-a\over r}-aC\log{r-a\over l}){\Big)},&(4.23)}$$
$$\li{T_{11}&={1\over96\p r^2(a-r)^2}{\Big(} 11a^2-12ar-12a^2Cr+18aCr^2
+(C^2-12CD)r^4+\cr
&+12(a-r)^2(\log {r-a\over r}-aC\log{r-a\over l}){\Big)}&(4.24)\cr}$$
From (4.23-24),
the ingoing and outgoing fluxes are
given by
$$\li{T_{uu}=T_{vv}&={1\over 384\pi r^4}(6a^2-8ar-24 Ca^2r+36aCr^2+\cr
&+2(C^2-12CD)r^4
+24(a-r)^2[(1-aC)\log (r-a)-\cr
&-\log r+aC\log l]).&(4.25)\cr}$$
For the values of $C$ and $D$ given by (4.22), the conditions of the
regularity of flux on the horizon calculated in the  Kruscal coordinates
(see Appendix of \cite{cf})
$$T_{vv}<\infty ,\ (r-a)^{-1}T_{uv}<\infty ,\ 
 (r-a)^{-2}T_{uu}<\infty \eqno(4.26)$$ 
are fulfilled. It is interesting to note that we could take (4.26) as
the regularity condition and obtain the same result (4.22) for $C$ and $D$.

With these values of the constants our system is in the same 2D Hartle-Hawking
vacuum state as in the null-dust case. This can be confirmed by performing a 
direct calculation of $<T_{uu}>$ and $<T_{vv}>$ \cite{bf}. From (4.25) we see 
that the system is in the
thermodynamical equilibrium, but the emission and
absorption fluxes at spatial infinity are negative, given by
 $-{5\over 192a^2}$. This property of SSG model can be understood from the fact
that unlike the null-dust case, we have a strongly interacting gas in a 1D box,
due to strong dilaton field at spatial infinity which couples to the 2D scalar
field. Consequently the effective potential energy density is negative enough 
to make the total energy density negative, and hence the negative flux.
As shown in the appendix A, the constants $C$ and $D$ which determine the
flux at infinity, do not depend on the regularization-dependent coefficient 
$c$, so that negative flux is not a regularization artefact.

Note that one can find states with positive flux at spatial infinity by 
choosing the values of constants $C$ and $D$ such that $C\sp 2 -12CD > 0$.
These will be also states of thermal equilibrium with temperature close to
the classical Hawking temperature. However,
the problem with such states is that the scalar curvature will diverge in the
vicinity of the horizon, so that their physical interpretation is not clear.

The values for the position of the horizon and the temperature in the 2D
Hartle-Hawking vacuum are respectively
$$r_h=a+\k\Big( -{23\over 4a}+{1\over 2a}\log {a\over l}\Big)\eqno(4.27)$$
and
$$4\p T_{H}={1\over a}\Big[1+\k({15\over 2a^2}+{5\over
2a^2}\log{L\over l})\Big]\,.\eqno(4.28)$$
The entropy
can be found by using the Wald method. 
For the lagrangian given by (4.2), from (3.16) we get
$$S={a^2\pi \over G}+{\p\k\over G}\left(-{15\over 2}+5\log {a\over l}\right)
\,. \eqno(4.29)$$ 
Note that the Euclidean method gives the same expression for the entropy
as (4.29) (see the Appendix B for details).
The first term in (4.29) is the
Bekenstein-Hawking entropy, while the first term is the quantum
correction. The quantum correction is of the same type as in the
previous case.
The energy of the system can be obtained from the second law of thermodynamics.
This gives
$$E=M-{5\k\over 4MG^2}\left({7\over 2}+\log {L\over l}\right)\, .\eqno(4.30)$$

\sect{5. Conclusions}

The 2D Hartle-Hawking boundary conditions which we have used for the SSG
model give negative energy density at spatial infinity, which is in contrast
with the 4D Hartle-Hawking vacuum state where the energy density is positive.
If one tries to impose the 4D HH boundary conditions, one quickly sees that
this is impossible in this model. The relation between the 4D energy-momentum
tensor and the corresponding 2D 
energy-momentum tensor for spherically symmetric models is given by 
$$T_{\mu\nu}= 4\pi r^2T _{\mu\nu}^{(4)}\ .\eqno(5.1)$$
From (4.23-4) and (5.1) it can be seen that the behaviour of 
$T _{\mu\nu}$ at spatial infinity is such that the corresponding 
$T_{\mu\nu}^{(4)}$ can not describe a 4D Hartle-Hawking vacuum since
 $T_{00}^{(4)}$ has the asymptotics 
$\sim {{\rm const}\over r^2}$ instead of $\sim$ const. This is not surprising
because the reflecting boundary conditions in 2D (gas in a 1D box) correspond
to the reflection of only the s-modes in 4D. For the
 4D Hartle-Hawking vacuum one needs boundary conditions corresponding to a
gas in a 3D box. Such boundary conditions can be implemented in a spherically
symmetric situation, as explicitly demonstrated in \cite{ahwy}; however,
the corresponding 2D effective action is clearly not the one given by (2.8).   
Including the extra local term coming from the action (2.9) does not improve 
the situation. Clearly a nontrivial modification is necessary, and
further work should be done.
 
The failure of the action (2.8) (which was obtained by the standard 
perturbative calculation) to describe the most general situations
of interest indicates that performing a spherical reduction first and then 
quantizing is not
equivalent to quantizing first and then performing a spherical reduction.
This is not surprising, given that in the first approach one neglects the
quantum effects of the angular modes. One way to see the effect of the
angular modes is to compare the results of 2D effective action approach to the
approach based on the equation (1.1). This requires a further work, because the
existing results in the later approach \cite{ahwy} use boundary conditions 
corresponding to a 4D HH state. For example, the function $m(r)$
is given by
$$\li{Km(r)&={r^3\over 3a^3}+{r^2\over a^2}+{3r\over a}-{13\over
3}\cr 
&+[22-120 (\xi-{1\over6})]{r-a\over r}+[30-240 (\xi-{1\over6})]
({r-a\over r})^2\cr
&-[11-120 (\xi-{1\over 6})]({r-a\over r})^3+4\log {r\over
a}+ C_0 &(5.2)\cr}$$
where $K=3840\pi$ and  $\xi Rf^2$ is an additional term in
the lagrangian density of the scalar field. Its asymptotic behaviour is 
$O(r\sp 3 )$, while the 2D HH vacuum requires the asymptotics $O(r)$.

In the 2D effective action approach one can compare the results for various 
models.
Frolov, Israel and Solodukhin in
\cite{fis} obtained for the 2D HH state
$$m(r)=-{7a\over 4r^2}+{1\over 2r}-{r\over 2a^2}-{1\over 2a}\log {r\over l}
\eqno(5.3)$$
Our result for the null-dust case is
$$m(r)=-{2\over r}-{r\over 2a^2}-{1\over 2a}\log {r\over l}\ .\eqno(5.4)$$
In the SSG model case 
the 2D Hartle-Hawking state gives
$$\li{m(r)=&{11a\over 4r^2}+{1\over 2r}+{5\over
2}r +\cr
&+{5\over 2a}\log {r\over l}-{6\over r}\log  {r\over l}+{3a\over r^2}
\log {r\over l}\,, &(5.5)\cr}
$$
while the states with positive Hawking flux (for some $C$) have
$$\li{m(r)&=-{11a\over 4r^2}+{-5+6aC\over 2r}-{1-6aC\over 2a^2}r
+(-{3a\over r^2}+{6\over r}-{5\over 2a})\log {r-a\over r}\cr
&+
({3a^2C\over r^2}-{6aC\over r}-{1-6aC\over 2a})\log {r-a\over l}\ ,&(5.6)\cr}$$
where (4.15) is taken. 
Note that the logarithmic divergence in (5.6)
for $r\to a$ is superficial, as the corresponding coefficient goes to 0
in this limit. 

As it can be seen, 
the form of the corrections of $m(r)$ in all given cases is similar; they
all diverge as $r\to 0$, which is to be expected. But it is interesting
to note  the large $r$ behaviour of $m(r)$: it goes as $r$ in the 2D case
while  $r^3$ in the
4D case. This means that the metric diverges for large $r$ in the 4D model 
\cite{ahwy},
while in 2D models it is finite. This gives some restrictions on the
applicability of 4D perturbative calculations. The
2D results which we noted above are all qualitatively similar. Unlike 
\cite{fis},
we obtained that the \Sch singularity stays at $r=0$ in the null-dust and
the SSG case. The usual 2D HH state has negative Hawking flux in the SSG case,
which is the consequence of the interaction between the dilaton an the matter
field. The positive Hawking flux states exist in the SSG model, but their
physical interpretation is not clear due to
curvature singularity at $r=a$. This is related to the
pathological behaviour of the flux for the freely falling observer at
the horizon in this case. One may think that these states may be related to a 
4D Unruh vacuum, but the relation $<T_{uu}> = <T_{vv}>$ does not allow this.
The quantum corrections for the entropy have
similar forms in various 2D models.

The SSG action was recently used in \cite{bh1} to obtain the quantum
correction to the Schwarzschild-de Sitter solution, which represents a 
non-asymptotically flat black hole. 
We  succeeded to put this action
in the local form introducing the two auxiliary
fields, which, we believe, will be of importance in future investigations of
this model.
It would be interesting to see how our
perturbative
analysis extends to the case of non-vacuum solutions which are time-dependent.
In this case it would be
useful to use the null-dust model as a zeroth-order approximation, since it
is exactly solvable \cite{mik}.

At the end of calculation our solutions contain two dimensional
constants: $l$ and $L$. This is the common case in the literature
\cite{fis,y,ahwy}: $l$ is short-distance cut-off,
of order of the Planck length, while $L$ describes the infrared
divergencies related to the radiated matter and is of order of the
magnitude of space. 

We obtained the entropy from the Wald approach in both cases
which we analysed. The
same value for the entropy is obtained from the Euclidean method. The
logarithmic form of the correction to the classical entropy is obtained, 
containing the cutoff
parameters $l$ and $L$. This  also occurs in other models of quantum 
black holes \cite{fis,su,s}. Let us note that the calculation of
entropy is more straightforward in 2D than in 4D.
By using the black hole thermodynamics laws we  determined the
corrected energy of the black hole. Our results imply that
$r_h \ne 2GM_{bh}$ in the quantum case. It would be interesting 
to calculate the ADM mass of the one-loop solution and compare it to the
value obtained from the thermodynamics.

\sect{Appendix A}

The general form of the action is
$$\li{S={1\over 4G}{\Big(}\int d^2x\sqrt{- g}&[ 
r^2R+2(\nabla r)^2+2]\cr
 -\kappa\int d^2x \sqrt{- g}&[ 2R(\psi -6\chi )+(\nabla \psi )^2-\cr
&-12 (\nabla \psi
)(\nabla \chi) -12{\psi (\nabla r)^2\over r^2}-12\g R\log r]{\Big)} \ ,
&(A.1)\cr } $$
where $\g = c/12$. In the previous analysis, $\g=1$. 

The perturbative solution of the equations of motion is
$$\li{m(r) =&{(36\g -25)a\over 4r^2}+{19-24\g +6aC\over 2r}-{1\over
2}(C^2-12CD)r -{5\over 2a}\log {r-a\over r}-\cr
&+{6\over r}\log  {r-a\over r}-{3a\over r^2}\log {r-a\over r}-\cr
&-{a(C^2-12CD)\over 2}\log {r-a\over l}
-{6aC\over r}\log {r-a\over l}+{3a^2C\over r^2}\log {r-a\over l}
\quad, &(A.2)\cr}
$$
$$\f (r)=\k (F(r)-F(L))\quad,\eqno(A.3)$$
 where $F(r)$ is given by
$$\li{F(r)=&{3(4\g -3)\over 4r^2}+{2+3aC\over
ar}+{1-6aC-a^2(C^2-12CD)\over 2a(r-a)}+ \cr
&+{5\over 2a}\log {r-a\over r}-{3\over r^2}\log {r-a\over r} +\cr
&+{(C^2-12CD)\over 2}\log {r-a\over l}+{3
aC\over r^2}\log {r-a\over l}
\quad. &(A.4)\cr}
$$
The values of the constants $C$ and $D $ do not depend on $\g$. Using
(A.2-4), it is straightforward to obtain the expressions for the
temperature, entropy, etc.

\sect{Appendix B}

In this appendix we will present the other derivation of the entropy of
the corrected Schwarzschild solution. We will calculate entropy using
the conical singularity method, developed in \cite{fis,s,fs}. 
In order to define thermodynamical properties of the system, one
considers the Euclidean theory. In that case, the free energy of the
system $F$ is proportional to the Euclidean  action. We will consider
the system at the arbitrary temperature $\bar T=(2\p\bar\b )^{-1}$.

In Euclidean theory, $\tau = it$, the metric (3.7) takes the form
$$ds^2=f(r)e^{2\f (r)} d\tau ^2+{1\over f(r)}dr^2\ ,\eqno(B.1) $$
where $\tau\in [0, 2\p\bar\b ]$.
If we define $\rho =\int {dr\over \sqrt{f(r)}}$, the metric (B.1) becomes
$$ds^2=g(\rho )d\tau ^2+d\r ^2\ 
,\eqno(B.2) $$
where $g(\r )=e^{2\f }f$. Near the horizon $\r \approx 0$, (B.2) may be
written in the form
$$ds^2={\r ^2\over \b _{H}^2} d\tau ^2+d\r ^2\ ,\eqno(B.3) $$
where ${1\over \b _{H}}= 2\pi T_{H}={1\over 2}e^{\f}f^{\prime}|_{r=r_{h}}$.
From (B.3) we see that for regular solution $\bar\b =\b _{H}$, meaning
 that the conical singularity in (B.3) is absent.
The metric (B.2) can be conformally 
related to the metric of the cone $C_\alpha$
\cite{fis,fs}:
$$ds^2=e^\s (dz^2+{1\over \b _{H}^2}z^2 d\tau ^2)=e^\sigma ds^2_{C_\a}\ .
\eqno(B.4)$$
Conformal factor $\s$ and coordinate $z$ 
can be found easily. The conical metrics in (B.4)
must be regularized at the tip of the cone. This is done using
\cite{fs} the regular metrics
$$ds^2_{\tilde{C_\a}}=u(z,a,\a)dz^2+{z^2\over \b _{H}^2} 
d\tau ^2\quad ,
\eqno(B.5)$$
where $a$ is the regularization parameter. The simplest choice for the
function $u$ is 
$$u={z^2+a^2\a^2\over z^2+a^2}.$$

Instead of the manifold $M_{\a}(C_{\a})$ we will consider the regularized
manifold $\tilde M_{\a}(\tilde C_{\a}).$ 
 On the regularized manifold our effective action (4.2) takes
the form
$$\li{\G=&-{1\over 4G}{\Big(} \int _{\tilde M_\a}d^2x\sqrt g[ 
r^2R+2(\nabla r)^2+2]-\cr
 -&\kappa\int d^2x \sqrt g [ 2R(\psi -6\chi )+(\nabla \psi )^2-\cr
 &-12 (\nabla \psi
)(\nabla \chi) -12{\psi (\nabla r)^2\over r^2}-12R\log r]{\Big)}-\cr
&-{1\over 2G}{\Big(}\int _{\pa \tilde {M_\a}}r^2kds -
\kappa
 \int _{\pa \tilde {M_\a}}(2\psi-12\chi-12\log r)kds\Big{)}\ .
&(B.6)\cr } $$
In the action (B.6) we added appropriate surface terms. $k$ is the
external curvature of the boundary of the manifold.
After the conformal transformation (B.4), from (B.6) we get
the action
$$\li{ \G
=&-{1\over 4G}{\Big(} \int _{\tilde C_\a}d^2x\sqrt {\tilde g}[ 
r^2\tilde R-r^2\tilde \nabla ^2\s +2(\tilde \nabla r)^2+2]\cr
 -&\kappa\int _{\tilde C_{a}}d^2x\sqrt
  {\tilde g} [ \tilde R\tilde \psi -12\tilde R\tilde \chi -2\s \tilde R+\s
\tilde \nabla ^2 \s 
+12 (\tilde \nabla \s
)(\tilde \nabla \chi) +\cr
&+ 12\tilde \chi \tilde \nabla ^2 \s -12\s \tilde \nabla ^2
\tilde \chi -
12\tilde R\log r+12 \log r\tilde \nabla ^2 \s]{\Big)}-\cr
&-{1\over 2G}\int
r^2(\tilde k +{1\over 2}\tilde n^{\m}\tilde \pa _{\m}\s)+{\k \over G}\int 
{\Big(}\tilde \psi \tilde k -3\tilde \psi \tilde n^a\pa _a \tilde\chi +\cr
&+{1\over 4}\s \tilde n^{\m}\tilde \pa _{\m}\s +{1\over 4}\tilde \psi \tilde
n^{\m} \pa _{\mu}\tilde \psi +
(\tilde k +{1\over 2}\tilde n^{\m}\pa _{\m}\s)(-\s -6\tilde \chi -6\log
r){\Big)}\ ,
&(B.7)\cr } $$
where $\tilde R={u^{\prime}\over zu^2}$ iz curvature of the
regularized cone $\tilde C_{\a}$, while $\tilde \psi =\psi -\s$ and
$\tilde \chi =\chi$. $\tilde k={1\over z\sqrt u}$ is
 the external curvature of the boundary of $\tilde C_\a.$ We now
take the limit $\tilde C_{\a}\to C_{\a}$. After that it is easy to
rewrite the effective action in terms of quantities defined on the manifold
$M_{\a}/ \S$ and $\S$, where $\S$ is the tip of the cone. Finally,
 we obtain  
$$\li{\G=&-{1\over 4G}{\Big(} \int _{ M_{\a}/\S}d^2x\sqrt g[ 
r^2\bar R+2(\nabla r)^2+2]-\cr
 &-\kappa\int _{M_{\a}/\S}d^2x \sqrt g [ 2\bar R(\bar \psi -6\bar \chi )
 +(\nabla \bar \psi )^2-\cr
 &-12 (\nabla \bar \psi
)(\nabla \bar \chi) -12{\bar \psi (\nabla r)^2\over r^2}-12\bar R\log r]{\Big)}
-\cr
&-{\pi \over G}(1-\a)r^2(\S)
+{2\pi \kappa \over G}(1-\a)(\bar \psi (\S)-6\bar \chi (\S)-6\log r(\S))-\cr
&-{1\over 2G}{\Big(}\int _{\pa M_\a}r^2\bar kds -\kappa
 \int _{\pa  M_\a}(2\bar \psi-12\bar \chi-12\log r)\bar
kds{\Big)}+O((1-\a )^2)
 \ ,&(B.8)\cr } $$
 where $\bar R$ is the regular part of curvature, $\bar \psi $ is the
solution of the equation (4.5) in the case $\a=1$. Also,  $\bar \chi =\chi $,
 while $\bar k=k|_{\a=1}$. 
Using the definition of entropy
$$S=(\a{\pa\over \pa\a}-1)\G|_{\a =1}\eqno(B.9)$$
 from (B.8) we easily get
 $$S={\p\over G}\left(r_h^2 -\k(2\psi_h -12\chi_h -12\log r_h)\right)\,
  \eqno(B.10)$$
up to the numerical coefficient. From (4.7-8), (4.27) and (B.10) we get 
the expressions for the entropy which are equal
to (4.29).

The similar calculation can be done for the null dust case, with the
same conclusion that the expression for the entropy is the same as (3.18).

\end{document}